\documentclass[12pt]{article}

\usepackage{amsmath,amsfonts,amssymb}
\usepackage[dvips]{graphics}
\usepackage[latin1]{inputenc}
\usepackage[dvips]{epsfig}
\newcommand{\MBB}[1]{\mathbb{#1}}
\newcommand{\MCA}[1]{\mathcal{#1}}
\newcommand{\MRM}[1]{\mathrm{#1}}
\newcommand{\MFR}[1]{\mathfrak{#1}}
\newcommand{\REF}[1]{(\ref{#1})}

\begin{document}

\begin{titlepage}
$\phantom{X}$\vspace{-20mm}\\ 
\noindent{$\phantom{X}$\hfill{\Large\sc MS-TPI-98-09}}\vspace{20mm}
\begin{center}
{\LARGE
On the Stability of the $O(N)$-Invariant \\[1mm]
and the Cubic-Invariant $3$-Dimensional \\[1mm] 
$N$-Component Renormalization Group Fixed \\[2mm]
Points in the Hierarchical Approximation}\\[10mm]
{\Large K.~Pinn, M.~Rehwald, Chr.~Wieczerkowski}\\[10mm]
{\large Institut f\"ur Theoretische Physik 1\\[1mm]
Westf\"alische Wilhelms-Universit\"at M\"unster\\[1.5mm]
Wilhelm-Klemm-Str.~9, D-48149 M\"unster}\\[10mm]
\end{center}
{\bf Abstract} We compute renormalization group fixed points and 
their spectrum in an ultralocal approximation. We study a case of 
two competing non-trivial fixed points for a three-dimensional real 
$N$-component field: the $O(N)$-invariant fixed point vs.~the 
cubic-invariant fixed point. We compute the critical value $N_{c}$ 
of the cubic $\phi^{4}$-perturbation at the $O(N)$-fixed point.
The $O(N)$ fixed point is stable under a cubic 
$\phi^{4}$-perturbation below $N_{c}$, above $N_{c}$ it 
is unstable. The critical value comes out as $2.219435<N_{c}<
2.219436$ in the ultralocal approximation. We also compute the 
critical value of $N$ at the cubic invariant fixed point. Within 
the accuracy of our computations, the two values coincide. 
\end{titlepage}

\section{Introduction}

Spin systems with an $N$-component real field variable, governed 
by a Landau-Ginzburg type Hamiltonian, are of central importance 
both in Euclidean quantum field theory and classical statistical 
mechanics. The basic model is one with a global $O(N)$ symmetry, 
the invariance under a simultaneous rotation of all the spins. A 
prototypical Hamiltonial for such a model, say on a lattice 
$\Lambda\subset\MBB{Z}^{D}$, is 
\begin{equation} 
H = 
\sum_{x\in\Lambda}\left\{
\frac12 
\sum_{a=1}^N 
\left(
\sum_{\mu=1}^{D} 
\partial_\mu \phi_a (x)^{2}
+\phi_a(x)^{2}
\right)
+ \lambda \left( \sum_{a=1}^N \phi_a(x)^2\right)^2 
\right\}. 
\label{1}
\end{equation} 
It is important to study the influence of perturbations which
explicitely break this symmetry to a smaller subgroup. 
For instance, in cubical crystals, one expects the spin interaction 
to react to the lattice structure. This suggests additional terms in
the Hamiltonian which are not rotation symmetric, but invariant under
the cubic group. The cubic group is composed of the permutations and
reflections of the $N$ components of the field. Such a
cubic invariant term is
\begin{equation}
\mu \sum_{x\in\Lambda}\sum_{a=1}^{N} \phi_a(x)^4.
\label{2}
\end{equation}
When (\ref{1}) is augmented by (\ref{2}), a competition of 
renormalization group fixed points sets in, about which of them 
determines the long distance behaviour of the model.

According to folklore, the $O(N)$ fixed point becomes unstable 
above a certain threshold value $N_c$. When this happens, the 
cubical fixed point dominates the long distance behaviour. 
The value of $N_c$ is still being debated.  Work on the 
$\varepsilon$-expansion suggests that $N_c > 3$. This has been 
challenged by the studies \cite{S1,S2,S3,S4,S5}, which suggest that 
$N_c < 3$.  Recent Monte Carlo work \cite{MM} by Caselle and 
Hasenbusch again indicates that $N_c$ should be very close to 3. 
Within their precision, this result is compatible with that of 
Kleinert and collaborators \cite{S3,S4,S5}.

To our knowledge, the problem has not yet been looked at in 
the framework of the block spin renormalization group 
\cite{Wilson/Kogut:1974}. To study the domains of attractions 
in a model with several competing fixed is a fundamental and
challenging problem as it underlies Wilson's explanation of 
universality. A notorious trouble in this business is
that these Hamiltonians in general depend on an infinite
number of interaction terms. To control flows with a large 
number of couplings is a very difficult task. 
Furthermore, when symmetries are reduced, the number of couplings 
tends to proliferate dramatically.
 
This paper contains a study of both the $O(N)$-invariant 
fixed point and the cubic fixed point in the framework
of the hierarchical or ultralocal approximation to Wilson's 
renormalization group. Here we restrict our attention to the case
of $D=3$ dimensions. In the hierarchical approximation, the 
effective Hamiltonians are restricted to a (non-standard) kinetic 
term plus local interactions. The local interactions are given by a 
potential, which is a function of $N$ variables. 

\section{Hierarchical Renormalization Group}

The hierarchical renormalization group is a block spin 
renormalization group for so called hierarchical spin models. 
They are spin models with a non-translation invariant kinetic
term designed to make the renormalization group local. 
The hierarchical model and the full model belong to different
universality classes. Nevertheless they are related. 
The hierarchical renormalization group is an ultralocal approximation
to a lattice block spin renormalization group. See 
\cite{Koch/Wittwer:1986,Pinn/Pordt/Wieczerkowski:1994} and 
references therein. 
Furthermore, the hierarchical model is presumably a zeroth approximation 
to the full model in a lattice derivative expansion. Hierarchical
results are to our experience reasonable approximations. 
However concerning their predictions about full models, high 
precision cannot be expected. For instance, $\nu$ comes out 
as $0.649$ at the $N=1$ Ising fixed point, as compared to 
$0.63$ from $\varepsilon$-expansion and Monte-Carlo methods. 
Hierarchical models are known to be extremely valuable in gaining 
qualitative 
information about the model under investigation and to 
prepare the ground for further renormalization group studies,
where non-local interactions are included.  

The hierarchical renormalization group for models with an $N$-component 
real scalar field is a theory of the non-linear integral transformation 
\begin{equation}
R_{\alpha,\beta,\gamma}(Z)(\psi)\;=\;
   \int\MRM{d}\mu_{\gamma}(\zeta)\;Z(\beta\psi+\zeta)^{\alpha},
\label{1.1}
\end{equation}
where $\alpha$, $\beta$, and $\gamma$ are real parameters, and where 
$\MRM{d}\mu_{\gamma}(\zeta)$ is the Gaussian measure on $\MBB{R}^{N}$
with mean zero and covariance $\gamma$ (times the unit matrix). Recall 
that its Fourier transform is
\begin{equation}
\int\MRM{d}\mu_{\gamma}(\zeta)\;
\MRM{e}^{\MRM{i}\zeta j}\;=\;
\MRM{e}^{-\frac{\gamma}{2}\,j^{2}}.
\label{1.2}
\end{equation}
We consider the renormalization group in the so called high temperature 
picture. See \cite{Pinn/Pordt/Wieczerkowski:1994} and references therein. 
Our parameter values in \REF{1.1} are
\begin{equation}
\alpha\,=\,2,\qquad
\beta\,=\,2^{-\frac{2+D}{2D}},\qquad
\gamma\,=\,1.
\label{1.3}
\end{equation}
(You may take this as the definition of the high temperature picture.)
Here $D$ is the dimension of the model. We restrict our attention to 
the most interesting case, when $D=3$. 

The transformation \REF{1.1} is the composition of three steps: 
taking the square, Gaussian convolution, and rescaling. The numerical 
computation of \REF{1.1} will be decomposed into these steps. 

\section{Algebraic Formulation}

We use the techniques described in 
\cite{Pinn/Pordt/Wieczerkowski:1994,Gottker:1996}.
Consider first the $O(N)$-invariant case. A general even $O(N)$-invariant 
Boltzmann factor can be written as a sum
\begin{equation}
Z(\phi)\;=\;
\sum_{n=0}^{\infty}
(\phi^{2})^{n}\;Z_{n},
\label{2.1}
\end{equation}
and is parametrized by real coefficients $Z_{n}$.
The transformation \REF{1.1} becomes the following non-linear transformation
in terms of the coordinates $(Z_{n})$:
\begin{equation}
R(Z)_{n}\;=\;
\beta^{2n}\;
\sum_{m=n}^{\infty} G_{n,m}(N,\gamma)\;
\sum_{l=0}^{m}Z_{l}\,Z_{m-l}.
\label{2.2}
\end{equation}
The coefficients $G_{n,m}(N,\gamma)$ are defined as follows. (We call
them structure coefficients.) Let 
$P_{m}(\phi)\,=\,(\phi^{2})^{m}$. The Gaussian convolution of it
is a linear combination
\begin{equation}
\int\MRM{d}\mu_{\gamma}(\zeta)\;
P_{m}(\psi+\zeta)\;=\;
\sum_{n=0}^{m}\;P_{n}(\psi)\;G_{n,m}(N,\gamma),
\label{2.3}
\end{equation}
with coefficients (essentially the number of contractions)
\begin{equation}
G_{n,m}(N,\gamma)\;=\;
\gamma^{m-n}
\prod_{l=1}^{m-n}\frac{(N+2(n+l-1))(n+l)}{l}.
\label{2.4}
\end{equation}
It is convenient to define $G_{n,m}(N,\gamma)\,=\,0$ for $n>m$. 

Analytical (and numerical) experience suggests a different 
normalization, namely 
$Z_{n}\,=\,X_{n}\;\rho^{n}/\sqrt{(2n)!}\,$ with a suitable 
constant $\rho$. For notational simplicity, we prefer to display 
the normalization \REF{2.1}.

For practical computations, one has to truncate the transformation 
\REF{2.2} to a finite number $K$ of non-zero coefficients. 
For $n\leq K$, the resulting transformation is
\begin{equation} 
R_{K}(Z)_{n}\;=\;
\beta^{2n}\sum_{m=n}^{K}\,G_{n,m}(N,\gamma)\;
\sum_{l=0}^{m}Z_{l}\,Z_{m-l}.
\label{2.5}
\end{equation}
In the high temperature picture, this truncation scheme is known to 
converge as $K\rightarrow\infty$. See 
\cite{Koch/Wittwer:1991} 
for a detailed analysis of the one component case. 

Renormalization group fixed points are approximated as stationary 
flows of the truncated transformation
\begin{equation}
R_{K}(Z^{\ast})_{n}\;=\;Z_{n}^{\ast}.
\label{2.6}
\end{equation}
The most interesting datum of a renormalization group fixed point
is its spectrum, from which one learns the behavior of the linearized 
flow in its vicinity. The linearized flows around a fixed point is
given by
\begin{equation}
{\rm D}R_{K}(Y)_{n}\;=\;
2\beta^{2n}\sum_{m=n}^{K}\,G_{n,m}(N,\gamma)\,
\sum_{l=0}^{m}Z_{l}^{\ast}\,Y_{m-l}.
\label{2.7}
\end{equation} 
The spectrum of $Z^{\ast}$ is the set of eigenvalues $\lambda^{(i)}$ 
of \REF{2.7}. The eigenvalues again are directly related to the 
critical exponents $\sigma^{(i)}$. See \cite{Wilson/Kogut:1974}. In 
our model, the relation is $\lambda^{(i)}=\alpha^{\sigma^{(i)}/D}$. 
(A peculiarity of hierarchical models is that the critical 
exponents are $L$-dependent.) 
The eigenvalue problem reads
\begin{equation}
{\rm D}R_{K}(Y^{(i)})_{n}\;=\;
\lambda^{(i)}\;
Y^{(i)}_{n}.
\label{2.8}
\end{equation}
The linearized renormalization group transformation \REF{2.2} can 
be brought to a manifestly symmetric form in the no-truncation 
limit, and is thus diagonalizable. In practice, \REF{2.8} turns 
out to be a very reliable way to determine the spectrum of 
renormalization group fixed points. 

\section{Numerical Results: $\boldsymbol{O(N)}$ Fixed Point}

We have searched for fixed points of the system of algebraic
equations \REF{2.5} with a Newton algorithm. The program was 
written in C++ using the data type {\tt long double} representation
for real numbers. In order to check roundoff errors, 
we compared the {\tt long double} with the 
(simple) {\tt double} representation and found no significant 
deviations. 
 
\begin{tiny}
\begin{table}[h]
\begin{center}
\begin{tabular}[c]{|l|l|l|l|l|}\hline
\multicolumn{1}{|c|}{$L_{\rm max}$} & \multicolumn{1}{c|}{$\lambda^{(1)}$} & 
\multicolumn{1}{c|}{$\lambda^{(2)}$} & \multicolumn{1}{c|}{$\lambda^{(3)}$} & 
\multicolumn{1}{c|}{$\lambda^{(4)}$} \\

\hline
10 & 1.33 & 0.73 & 0.31 & 0.10  \\
12 & 1.36 & 0.79 & 0.38 & 0.15 \\
14 & 1.37 & 0.83 & 0.438 & 0.19 \\
16 & 1.384 & 0.849 & 0.469 & 0.22 \\
18 & 1.3854 & 0.854 & 0.4846 & 0.25 \\
20 & 1.3856 & 0.8560 & 0.48997 & 0.262 \\
22 & 1.38573 & 0.8562 & 0.49143 & 0.267\\
24 & 1.385742 & 0.85633 & 0.49175 & 0.268 \\
26 & 1.3857434 & 0.8563400 & 0.491812 &0.2691 \\
28 & 1.38574348 & 0.8563408 & 0.4918214 & 0.26923 \\
30 & 1.385743489 & 0.85634089 & 0.49182258 & 0.269247 \\
32 & 1.38574349013 & 0.8563409057 & 0.491822725 & 0.2692491 \\
34 & 1.385743490193 & 0.85634090651 & 0.491822739 & 0.26924937  \\
36 & 1.3857434901972 & 0.856340906576 & 0.4918227412 & 0.2692493996 \\
\hline
\end{tabular}
\end{center}
\caption{Spectrum of $O(2)_{\displaystyle 3}$ fixed point}
\label{MM.1}
\end{table}
\end{tiny}
\begin{figure}[htbp]
\begin{center}
\begingroup\makeatletter\ifx\SetFigFont\undefined%
\gdef\SetFigFont#1#2#3{%
\reset@font\fontsize{#1}{#2pt}%
\fontfamily{#3}
\selectfont}%
\fi\endgroup%
\begin{picture}(0,0)%
\includegraphics{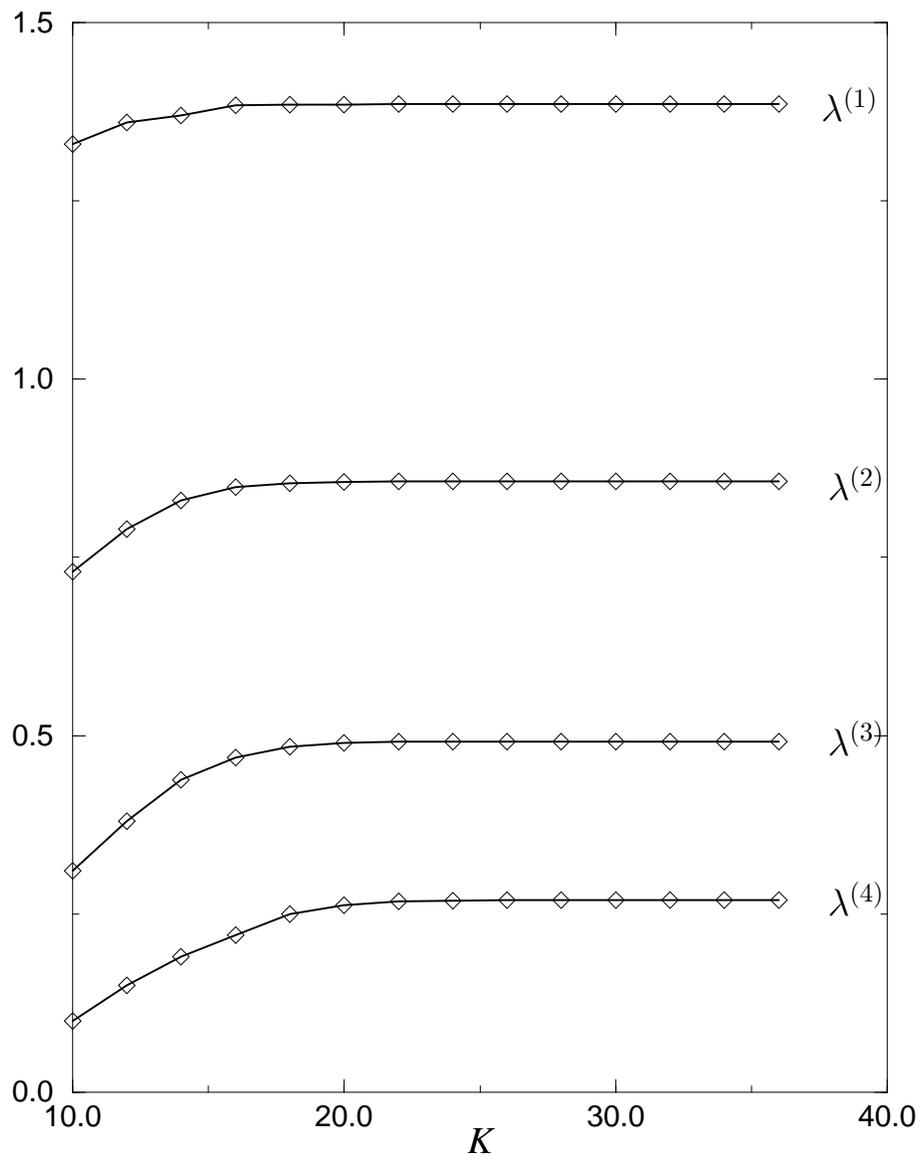}
\end{picture}%
\setlength{\unitlength}{0.24pt}%
\begingroup\makeatletter\ifx\SetFigFont\undefined%
\gdef\SetFigFont#1#2#3{%
\reset@font\fontsize{#1}{#2pt}%
\fontfamily{#3}
\selectfont}%
\fi\endgroup%
\begin{picture}(1427,1855)(269,313)
\put(1533,1988){\makebox(0,0)[lb]{\smash{\SetFigFont{14.4}{24.0}{cmr}
$\lambda^{(1)}$}}}
\put(1543,1391){\makebox(0,0)[lb]{\smash{\SetFigFont{14.4}{24.0}{cmr}
$\lambda^{(2)}$}}}
\put(1543,988){\makebox(0,0)[lb]{\smash{\SetFigFont{14.4}{24.0}{cmr}
$\lambda^{(3)}$}}}
\put(1543,735){\makebox(0,0)[lb]{\smash{\SetFigFont{14.4}{24.0}{cmr}
$\lambda^{(4)}$}}}
\end{picture}
\end{center}
\caption{Spectrum of $O(2)_{\displaystyle 3}$ fixed point as function 
of truncation order $K$}
\end{figure}

Table \ref{MM.1} shows the effect of truncation on the first few
eigenvalues of the non-trivial $O(N)$ fixed point with $N=2$
components in $D=3$ dimensions. (A trivial ``volume eigenvalue''
$\lambda^{(0)}=2$ has been omitted.) We see a rapid increase of
accuracy with the number of couplings. To get the same absolute
precision, the smaller the eigenvalue of interest is, the bigger the
number of couplings has to be chosen. This will be important in the
following.

The first eigenvalue $\lambda^{(1)}$ belongs to the critical exponent 
$\nu=1/\sigma^{(1)}$, or $\nu\,=\,\log(\alpha)/D\log(\lambda)$. 
The numerical value of $\nu$ is thus
\begin{equation} 
\nu\,=\,0.7082249.
\label{3.1}
\end{equation}
With twenty couplings, the first eigenvalue comes out with an 
accuracy of more than three digits, with thirty couplings of 
more than nine digits. 

A computer assisted proof for the one component case was developed 
by \cite{Koch/Wittwer:1991}. From it, one gets two sided bounds on 
the critical exponents with arbitrary precision. It confirms 
the accuracy of our calculation by a comparison at $N=1$. 

\begin{small}
\begin{table}
\begin{center}
\begin{tabular}[c]{|r|l|l|l|l|}\hline

\multicolumn{1}{|c|}{$N$} & \multicolumn{1}{c|}{$\lambda^{(1)}$} & 
\multicolumn{1}{c|}{$\lambda^{(2)}$} & 
\multicolumn{1}{c|}{$\lambda^{(3)}$} & \multicolumn{1}{c|}{$\lambda^{(4)}$} \\
\hline
-2.2 & 1.59824229 & 0.848707935 & 0.435463969 & 0.213609627\\
-2.0 & 1.58740105 & 0.849947302 & 0.437978578 & 0.215734559\\
-1.4 & 1.55436959 & 0.853424303 & 0.445855470 & 0.222533463\\
-0.6 & 1.50993452 & 0.857177559 & 0.457002524 & 0.232579287\\
-0.2 & 1.48805474 & 0.858501342 & 0.462742151 & 0.237983202\\
0.2 & 1.466769838 & 0.859352758 & 0.468493388 & 0.243587863\\
0.6 & 1.446379983 & 0.859668728 & 0.474160525 & 0.249329758\\
1.0 & 1.427172478 & 0.859411649 & 0.479637300 & 0.255127880\\
1.4 & 1.409390602 & 0.858578978 & 0.484816590 & 0.260888802\\
1.6 & 1.401091348 & 0.857956757 & 0.487263833 & 0.263724669 \\
2.0 & 1.385743489 & 0.856340897 & 0.491822589 & 0.269247586 \\ 
2.6 & 1.365884074 & 0.853161847 & 0.497720103 & 0.277002053\\
3.0 & 1.354668292 & 0.850694212 & 0.500989433 & 0.281718313\\
3.2 & 1.349623859 & 0.849396273 & 0.502426577 & 0.283920286 \\
\hline
\end{tabular}
\end{center}
\caption{Spectrum of $O(N)_{\displaystyle 3}$ fixed point
at $K=30$}
\label{MM.2}
\end{table}
\end{small}

\begin{figure}[htbp]
\begin{center}
\begingroup\makeatletter\ifx\SetFigFont\undefined%
\gdef\SetFigFont#1#2#3{%
\reset@font\fontsize{#1}{#2pt}%
\fontfamily{#3}
\selectfont}%
\fi\endgroup%
\begin{picture}(0,0)%
\includegraphics{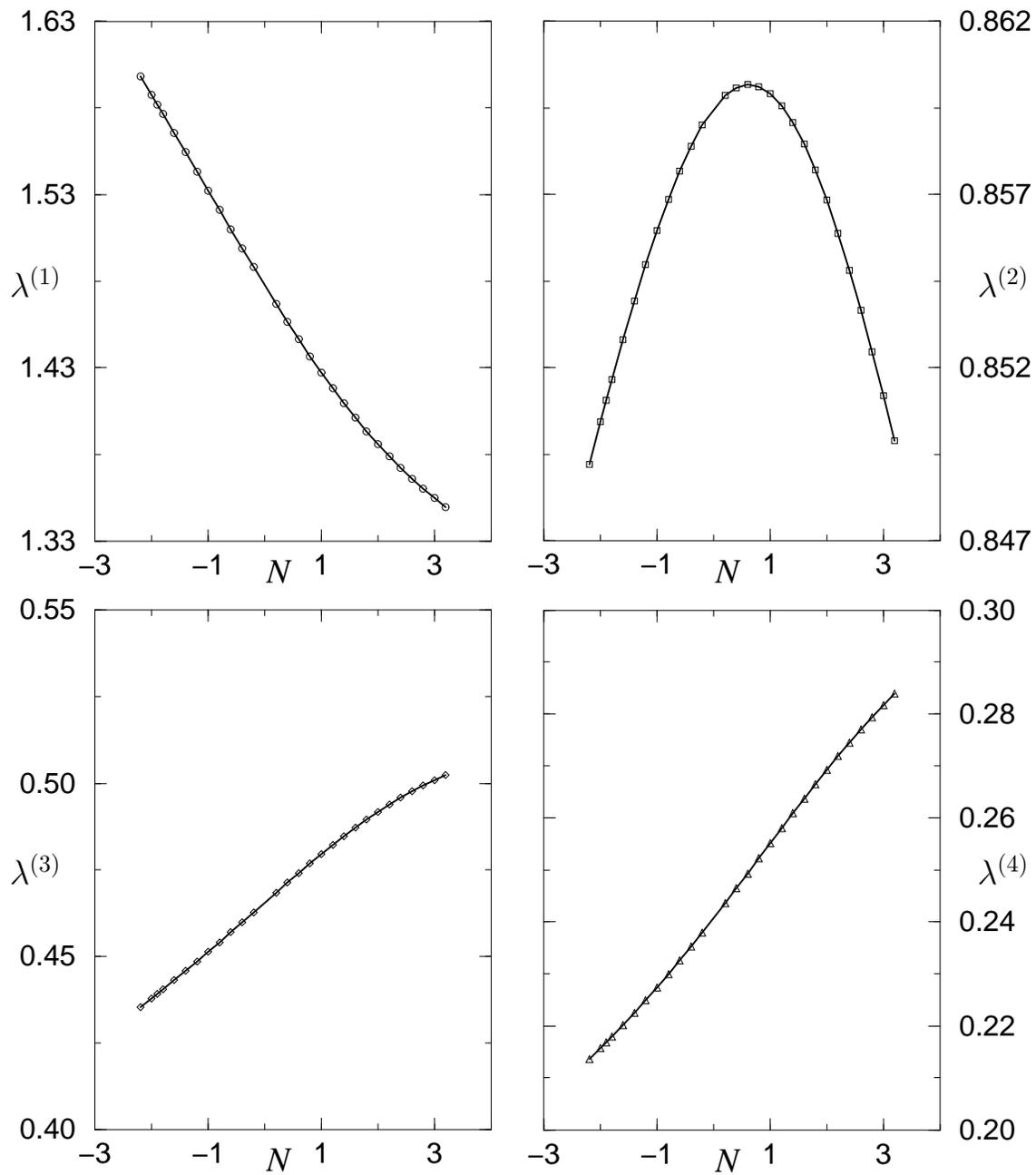}
\end{picture}%
\setlength{\unitlength}{0.24pt}%
\begingroup\makeatletter\ifx\SetFigFont\undefined%
\gdef\SetFigFont#1#2#3{%
\reset@font\fontsize{#1}{#2pt}%
\fontfamily{#3}
\selectfont}%
\fi\endgroup%
\begin{picture}(1770,2116)(138,217)
\put(221,764){\makebox(0,0)[rb]{\smash{\SetFigFont{14.4}{24.0}{cmr}
$\lambda^{(3)}$}}}
\put(221,1784){\makebox(0,0)[rb]{\smash{\SetFigFont{14.4}{24.0}{cmr}
$\lambda^{(1)}$}}}
\put(1793,764){\makebox(0,0)[lb]{\smash{\SetFigFont{14.4}{24.0}{cmr}
$\lambda^{(4)}$}}}
\put(1793,1784){\makebox(0,0)[lb]{\smash{\SetFigFont{14.4}{24.0}{cmr}
$\lambda^{(2)}$}}}
\end{picture}
\end{center}
\caption{Spectrum of $O(N)_{\displaystyle 3}$ fixed point
at $K=30$}
\end{figure}

Table \ref{MM.2} shows the spectrum of the non-trivial three 
dimensional $O(N)$-fixed point as a function of the number of
components $N$ at truncation order $K=30$.

Analytic continuation in $N$ is naturally possible for the system
of algebraic equations \REF{2.5}, since the $N$-dependence is encoded 
entirely in the structure constants \REF{2.3}. These depend 
polynomially on $N$. To maintain the same 
precision as $N$ increases, one has to increase $K$. Up 
to $N=3$, the truncation order $K=30$ suffices. 

We rediscover among other things the well known result that the 
theory becomes trivial at $N=-2$, i.~e.~, $\nu=0.5$.

\section{Cubic Invariance}

Cubic symmetry is the following. Consider an $N$-component model.
(The local spin takes values in $\MBB{R}^{N}$.) The cubic symmetry 
is the finite group of transformations, consisting of rotations
and reflections, which leaves invariant the cube $[-1,1]^{N}$.

We will restrict our attention to even Boltzmann factors, with
the property that
\begin{equation}
Z((-1)^{\sigma_{1}}\phi_{1},\ldots,(-1)^{\sigma_{N}}\phi_{N})\,=\,
Z(\phi_{1},\ldots,\phi_{N}).
\label{4.1}
\end{equation}
Such Boltzmann factors are functions of 
$\phi_{1}^{2},\ldots,\phi_{N}^{2}$.
To be cubic invariant, they have to be in addition symmetric 
functions of $\phi_{1}^{2},\ldots,\phi_{N}^{2}$. An even cubic 
invariant Boltzmann factor is thus a function \REF{4.1}, which in 
addition satisfies
\begin{equation}
Z(\phi_{\pi(1)},\ldots,\phi_{\pi(N)})\,=\;
Z(\phi_{1},\ldots,\phi_{N})
\label{4.2}
\end{equation} 
for all permutations $\pi\in\MFR{S}_{N}$. We begin with the 
simplest case, the study of cubic perturbations of the $O(N)$-invariant
fixed point. 

\subsection{Cubic Perturbations of the $\boldsymbol{O(N)}$ Fixed Point}

We can extend the polynomial basis $P_{n}(\phi)$ of $O(N)$-invariants
as follows. Let $\underline{n}=(n_{o},n_{c})\in\MBB{N}\times\{0,1\}$ 
and define
\begin{equation}
P_{\underline{n}}(\phi)\,=\,
\left(\sum_{a=1}^{N}\phi_{a}^2\right)^{n_{o}}\;
\left(\sum_{a=1}^{N}\phi_{a}^4\right)^{n_{c}}.
\label{4.3}
\end{equation}
These symmetric functions do not close under multiplication. They do 
however close under Gaussian convolution. They satisfy 
\begin{equation}
\int\MRM{d}\mu(\zeta)\;P_{\underline{m}}(\psi+\zeta)\;=\;
\sum_{\underline{n}}
P_{\underline{n}}(\phi)\;
G_{\underline{n},\underline{m}}(N,\gamma),
\label{4.4}
\end{equation} 
with structure coefficients
\begin{eqnarray}
G_{(n,0),(m,1)}(N,\gamma)&=&
  N\;\,G_{0,2}(1,\gamma)\;\,G_{n,m}(N+4,\gamma)
\nonumber\\&\phantom{=}&
  +G_{1,2}(1,\gamma)\;\,G_{n-1,m}(N+6,\gamma),
\label{4.5}\\
G_{(n,1),(m,1)}(N,\gamma)&=&
   G_{2,2}(1,\gamma)\;\,G_{n,m}(N+8,\gamma),
\label{4.6}\\
G_{(n,0),(m,0)}(N,\gamma)&=&
   G_{n,m}(N,\gamma),
\label{4.7}
\end{eqnarray}
where $G_{n,m}(N,\gamma)$ is as in \REF{2.4}. The linearized 
renormalization group at the $O(N)$ fixed point with cubic 
$\phi^{4}$-perturbations complicates to 
\begin{gather}
DR_{K}(Y)_{(n_{o},n_{c})}\;=\;
\nonumber\\
2\,\beta^{2n_{o}+4n_{c}}\,
\sum_{m_{o}=n_{o}}^{K}\sum_{m_{c}=0}^{1}\,
G_{(n_{o},n_{c}),(m_{o},m_{c})}(N,\gamma)\,
\sum_{l_{o}=0}^{m_{o}}Z_{(l_{o},0)}^{\ast}\,Y_{(m_{o}-l_{o},m_{c})}.
\label{4.8}
\end{gather}
The $O(N)$-invariants, defined by $n_{c}=0$, form an invariant
subspace. The cubic invariant eigenvectors generally
have non-vanishing $O(N)$-components. The cubic eigenvalues will
be denoted by $\kappa^{(i)}$. We order them according to their 
degree of relevance.

\begin{table}[htb]
\begin{center}
\begin{tabular}[c]{|l|l|l|l|l|l|l|l|}\hline
\multicolumn{1}{|c|}{$N$} & \multicolumn{1}{c|}{$\lambda^{(1)}$} & 
\multicolumn{1}{c|}{$\kappa^{(1)}$} & \multicolumn{1}{c|}{$\lambda^{(2)}$} 
& \multicolumn{1}{c|}{$\kappa^{(3)}$} & \multicolumn{1}{c|}{$\lambda^{(3)}$} \\
\hline
1.80 &  1.39321 & 0.98503 & 0.85721 & 0.54062 & 0.48960  \\
1.90 &  1.38942 & 0.98865 & 0.85679 & 0.54359 & 0.49072  \\
2.00 &  1.38574 & 0.99224 & 0.85634 & 0.54656 & 0.49182  \\ 
2.10 &  1.38217 & 0.99580 & 0.85588 & 0.54950 & 0.49288   \\
2.15 &  1.38042 & 0.99757 & 0.85562 & 0.55097 & 0.49340 \\
2.20 &  1.37870 & 0.99932 & 0.85537 & 0.55243 & 0.49391 \\
2.25 &  1.37701 & 1.00107 & 0.85511 & 0.55389 & 0.49442   \\
2.30 &  1.37534 & 1.00280 & 0.85484 & 0.55534 & 0.49491    \\
2.40 &  1.37209 & 1.00624 & 0.85430 & 0.55823 & 0.49588  \\
2.50 &  1.36893 & 1.00965 & 0.85374 & 0.56109 & 0.49681 \\
\hline
\end{tabular}
\end{center}
\caption{Cubic spectrum of $O(N)_{\displaystyle 3}$ fixed point at 
$K=30$}
\label{MM.3}
\end{table}
\begin{figure}[htb]
\begin{center}
\begingroup\makeatletter\ifx\SetFigFont\undefined%
\gdef\SetFigFont#1#2#3{%
\reset@font\fontsize{#1}{#2pt}%
\fontfamily{#3}
\selectfont}%
\fi\endgroup%
\begin{picture}(0,0)%
\includegraphics{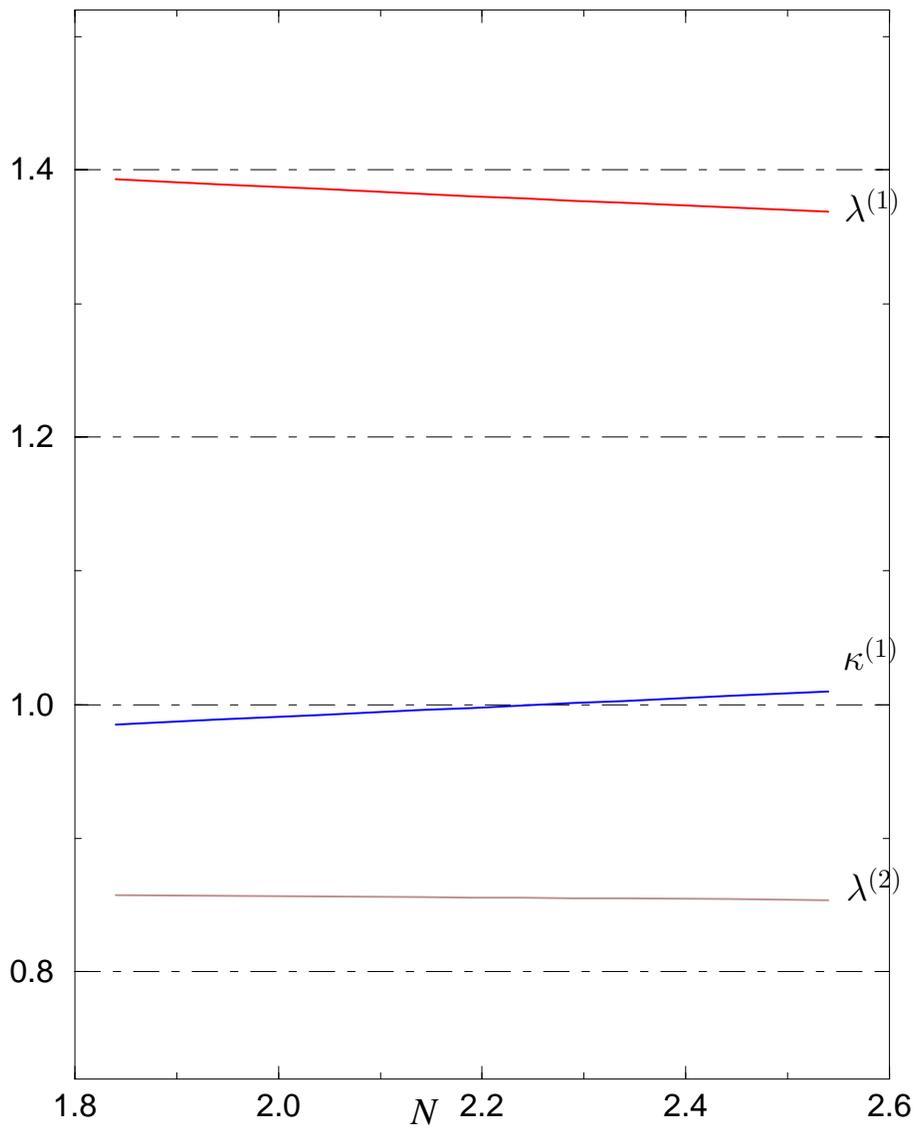}
\end{picture}%
\setlength{\unitlength}{0.24pt}%
\begingroup\makeatletter\ifx\SetFigFont\undefined%
\gdef\SetFigFont#1#2#3{%
\reset@font\fontsize{#1}{#2pt}%
\fontfamily{#3}
\selectfont}%
\fi\endgroup%
\begin{picture}(1424,1796)(261,338)
\put(1566,1808){\makebox(0,0)[lb]{\smash{\SetFigFont{14.4}{24.0}{cmr}
$\lambda^{(1)}$}}}
\put(1564,1101){\makebox(0,0)[lb]{\smash{\SetFigFont{14.4}{24.0}{cmr}
$\kappa^{(1)}$}}}
\put(1569,737){\makebox(0,0)[lb]{\smash{\SetFigFont{14.4}{24.0}{cmr}
$\lambda^{(2)}$}}}
\end{picture}
\end{center}
 \caption{Visualization of Table \REF{MM.3}} 
\end{figure}

Table \ref{MM.3} shows the $O(N)$-invariant and the cubic invariant
spectrum at the $O(N)$-invariant non-trivial fixed point in three
dimension as a function of the number of components $N$. The leading
cubic eigenvector is -- in $\varepsilon$-expansion a deformation of --
a cubic $\phi^{4}$-vertex. Here we restrict our attention to
eigenvectors of the $\phi^{4}$-type times powers of $\phi^{2}$.  (The
subleading eigenvalue $\kappa^{(2)}$ belonging to the cubic 
$\phi^6$-interaction is not displayed here.)

The largest cubic eigenvector, $\kappa^{(1)}$, becomes one at a
critical value $N_{c}$ of $N$. We learn from Table \ref{MM.3} that
$2.20<N_{c}<2.25$. A closer look at the vicinity of $N_{c}$ yields the
following.  Table \REF{tab:MM99} shows that the value of $N_{c}$ is
located inbetween 
\begin{equation}
2.219435<N_{c}<2.219436.
\end{equation}
(Numerical errors are
negligible.) We learn furthermore the important lesson that the
$N$-dependence of $\kappa^{(1)}$ is rather weak between the two and
three component models.

\begin{table}[htbp]
  \begin{center}
    \begin{tabular}[c]{|l|l|l|}\hline
 \multicolumn{1}{|c|}{$N$} & $K$ &  
\multicolumn{1}{c|}{$\kappa^{(1)}$} \\
\hline
2.219    & \footnotesize $\begin{array}{l} 30\\ 40\end{array}$ & 
\footnotesize$\begin{array}{l}0.99998479312540\\0.99998479665524\end{array}$\\
\hline
2.2194   & \footnotesize$\begin{array}{l}30\\40\end{array}$ & 
\footnotesize$\begin{array}{l} 0.99999876426610\\0.99999876780069\end{array}$\\
\hline
2.21943  & \footnotesize$\begin{array}{l}30\\40\end{array}$ & 
\footnotesize$\begin{array}{l}0.99999981395124\\0.99999981748619\end{array}$\\
\hline
2.219435 & \footnotesize$\begin{array}{l}30\\40\\50\end{array}$ &
\footnotesize$\begin{array}{l}0.99999998889842\\0.99999999243344\\
0.99999999243345\end{array}$ \\
\hline
2.219536 & \footnotesize$\begin{array}{l}30\\40\\50\end{array}$ & 
\footnotesize$\begin{array}{l} 1.00000002222169\\1.00000002575671\\
1.00000002575672\end{array}$\\
\hline
2.21944  & \footnotesize$\begin{array}{l}30\\40\end{array}$ & 
\footnotesize$\begin{array}{l} 1.00000016384551\\ 1.00000016738059 
\end{array}$\\
\hline
2.2195   & \footnotesize$\begin{array}{l}30\\40\end{array}$ & 
\footnotesize$\begin{array}{l} 1.00000226320303\\1.00000226673882\end{array}$\\
\hline
2.220    & \footnotesize$\begin{array}{l}30\\40\end{array}$ & 
\footnotesize$\begin{array}{l}1.00001973231681\\1.00001973585855\end{array}$\\
\hline

    \end{tabular}
    \caption{$N_c$ at the $O(N)$ fixed point}
    \label{tab:MM99}
  \end{center}
\end{table}

\section{Cubic Invariant Fixed Point}

Besides the $O(N)$-invariant fixed point, we find a cubic 
invariant fixed point. Again we restrict our attention to
the case of three dimensions. 
The cubic perturbations \REF{4.3} have to be enlarged to 
a generating system of cubic invariant polynomials. We 
have investigated several possibilities. 

\subsection{Lifted Representation}

The first possibility uses an over complete system. 
Let $\underline{n}=(n_{1},n_{2},n_{3},\ldots)
\in\MBB{N}\times\MBB{N}\times\MBB{N}\times\cdots$. Define 
\begin{equation}
P_{\underline{n}}(\phi)\;=\;
\left(\sum_{a=1}^{N}\phi_{a}^2\right)^{n_{1}}\;
\left(\sum_{a=1}^{N}\phi_{a}^4\right)^{n_{2}}\;
\left(\sum_{a=1}^{N}\phi_{a}^6\right)^{n_{3}}\cdots.
\label{5.1}
\end{equation}
We represent our fixed point by coordinates $Z_{\underline{n}}$.
With each collection is associated a function
\begin{equation}
Z(\phi)\;=\;\sum_{\|\underline{n}\|\leq {K}}\;
P_{\underline{n}}(\phi)\;
Z_{\underline{n}}.
\label{5.2}
\end{equation}


To define a suitable truncation, we introduce the norm
$  \|n\| := n_1 + 2 n_2 + 3n_3 \dots$. In other words, we 
truncate the model to a maximal power of fields. 
The summation is restricted to the finite subset 
of $\MBB{N}^{\infty}$ given by $\|n\| \leq K$. 

Unfortunately, the 
functions \REF{5.1} are not linearly independent. Moreover,
the linear dependencies vary with $N$. Therefore, the 
representation \REF{5.2} is not unique. 
As an illustration, the situation for $N=2$ is studied 
in detail in the Appendix. 

As we intend to use
the over complete representation, we have to specify a lift
of the renormalization group. We do this as follows. 
We have that
\begin{equation}
P_{\underline{n}}(\phi)\;
P_{\underline{m}}(\phi)\;=\;
P_{\underline{n}+\underline{m}}(\phi).
\label{5.3}
\end{equation}
The Gaussian convolution can be written in the form
\begin{equation}
\int\MRM{d}\mu_{\gamma}(\zeta)\;
P_{\underline{m}}(\psi+\zeta)\;=\;
\sum_{\|\underline{n}\|\leq{K}}\;
P_{\underline{n}}(\psi)\;
G_{\underline{n},\underline{m}}(N,\gamma),
\label{5.4}
\end{equation}
where the coefficients are the following. We compute
\begin{equation}
\sum_{a=1}^{D}\frac{\partial^{2}}{\partial\phi_{a}^{2}}
P_{\underline{m}}(\phi)\;=\;
\sum_{\underline{n}\leq\underline{m}}
P_{\underline{n}}(\phi)\;
\MCA{G}_{\underline{n},\underline{m}}
\label{5.5}
\end{equation}
with structure coefficients given by
\begin{alignat}{2}
\MCA{G}_{\underline{n},\underline{m}}&=
4ab\,m_{a}\,m_{b} 
\qquad &&
\exists~ a\ne b:
\begin{cases}
m_{a}-1=n_{a}\\
m_{b}-1=n_{b}\\
m_{a+b-1}+1=n_{a+b-1}\\
\forall~ c\ne a,b:~m_{c}=n_{c}
\end{cases}
\label{5.6}\\
\MCA{G}_{\underline{n},\underline{m}}&=
4a^{2}\,m_{a}\,(m_{a}-1)
&&
\exists~ a\ne 1:
\begin{cases}
m_{a}-2=n_{a}\\
m_{2a-1}+1=n_{2a-1}\\
\forall~ b\ne a:~m_{b}=n_{b}
\end{cases}
\label{5.7}\\
\MCA{G}_{\underline{n},\underline{m}}&=
2a\,(2a-1)\,m_{a}
&&
\exists~ a\ne 1:
\begin{cases}
m_{a}-1=n_{a}\\
m_{a-1}+1=n_{a-1}\\
\forall~ b\ne a:~m_{b}=n_{b}
\end{cases}
\label{5.8}\\
\MCA{G}_{\underline{n},\underline{m}}&=
2\bigl(2(m_{1}-1)+N\bigr)m_{1}
&&
\phantom{\exists~ a\ne 1:}
\begin{cases}
m_{1}-1=n_{1}\\
\forall~ a\ne 1:~m_{a}=n_{a}
\end{cases}
\label{5.9}
\end{alignat}
We then compute
the Gaussian integral as the matrix exponential thereof. The 
result is
\begin{equation}
G_{\underline{n},\underline{m}}(N,\gamma)\;=\;
\exp\left(\frac{\gamma}{2}\;
\MCA{G}\right)_{\underline{n},\underline{m}}.
\label{5.10}
\end{equation}
The matrix $\MCA{G}$ becomes upper triangular, when the 
couplings are sorted according to their total power of fields
$\|\underline{n}\|$.
(When the renormalization group is truncated to a finite power of
fields, the Gaussian convolution alone does not generate higher 
powers.)
Furthermore, the matrix $\MCA{G}$ becomes nilpotent, and the 
matrix exponential becomes a finite sum. Our first system of 
equations is 
\begin{equation}
R_{K}(Z)_{\underline{n}}\;=\;
\beta^{2\,\|\underline{n}\|}\;
\sum_{\|\underline{m}\|\leq{K}}
G_{\underline{n},\underline{m}}(N,\gamma)\;
\sum_{\underline{l}+\underline{k}=\underline{m}}
Z_{\underline{l}}\;Z_{\underline{k}}~~.
\label{5.12}
\end{equation}
This system of equations defines a lifting of the renormalization
group. Every fixed point of \REF{5.12} becomes through \REF{5.2}
a fixed point of the original hierarchical renormalization group. 
Their spectra of eigenvalues coincide (in the no-truncation limit), 
but some eigenvalues become degenerate. The eigenvalue problem is 
completely analogous to the $O(N)$-invariant case. We do not 
write down the equations here.

We can use \REF{5.12} to determine the cubic fixed point for any 
value of $N$. The reason is again that $N$ enters polynomially in 
the structure coefficients \REF{5.10}. The drawback of this 
representation is that the number of couplings increases very 
fast with the order of truncation. Table \REF{MM.4} shows this 
number of couplings with $\|n\|\leq K$ for 
$K=1\ldots 29$.

\begin{table}
\begin{center}
\begin{tabular}[c]{r|rp{5mm}r|rp{5mm}r|r}
\multicolumn{1}{c|}{$K$} &\#coups &&
\multicolumn{1}{c|}{$K$} &\#coups &&
\multicolumn{1}{c|}{$K$} &\#coups \\
\cline{1-2}\cline{4-5}\cline{7-8}
1 & 2 &&    10 & 139 &&     20 & 2714\\                        
2 & 3 &&    11 & 195 &&     21 & 3506\\                   
2 & 4 &&    12 & 272 &&     22 & 4508\\                   
3 & 7 &&    13 & 373 &&     23 & 5763\\                   
4 & 12 &&   14 & 508 &&     24 & 7338\\                    
5 & 19 &&   15 & 684 &&     25 & 9296\\                    
6 & 30 &&   16 & 915 &&     26 & 11732\\                   
7 & 45 &&   17 & 1212 &&    27 & 14742\\                   
8 & 67 &&   18 & 1597 &&    28 & 18460\\                   
9 & 97 &&   19 & 2087 &&    29 & 23025%
\end{tabular}
\end{center}
\label{MM.4}
\caption{Number of couplings in the lifted system for given truncation
$K$.}
\end{table}

We looked at the system up to $K=18$. In this model one 
has to compute $1597$ couplings. To get an idea of the achieved precision, 
we did one run with $K=19$, which means $2087$ couplings. 
(Has anyone ever computed a renormalization group with more unknowns?)

\begin{table}[htbp]
\begin{center}
  \begin{tabular}[c]{|l|r|r|r|r|r|r|}\hline
\multicolumn{1}{|c|}{N} & \multicolumn{1}{c|}{$\mu^{(1)}$} & 
\multicolumn{1}{c|}{$\mu^{(2)}$} & \multicolumn{1}{c|}{$\mu^{(3)}$} & 
\multicolumn{1}{c|}{$\mu^{(4)}$} & \multicolumn{1}{c|}{$\mu^{(5)}$} & 
\multicolumn{1}{c|}{$\mu^{(5')}$} \\ 
\hline
2.00  &  1.4337  &  1.0188  &  0.8540  &  0.5968  &  0.4887  &  0.4697\\
2.10  &  1.3988  &  1.0066  &  0.8551  &  0.5712  &  0.5367  &  0.4759\\
2.15  &  1.3857  &  1.0034  &  0.8540  &  0.5717  &  0.5545  &  0.4811\\
2.20  &  1.3790  &  1.0013  &  0.8531  &  0.5963  &  0.5500  &  0.4840\\
2.25  &  1.3765  &  0.9996  &  0.8525  &  0.6163  &  0.5481  &  0.4844\\
2.30  &  1.3758  &  0.9980  &  0.8521  &  0.6309  &  0.5480  &  0.4836\\
2.40  &  1.3761  &  0.9952  &  0.8515  &  0.6498  &  0.5491  &  0.4809\\
2.50  &  1.3766  &  0.9926  &  0.8508  &  0.6606  &  0.5498  &  0.4783\\
2.70  &  1.3765  &  0.9880  &  0.8490  &  0.6704  &  0.5489  &  0.4740\\
3.00  &  1.3735  &  0.9813  &  0.8451  &  0.6730  &  0.5442  &  0.4685\\
\hline
\end{tabular}
\end{center}
\label{MM.5}
\caption{Cubic spectrum at $K=18$.}
\end{table}

As in the $O(N)$ case we recognize one eigenvalue to become marginal at a
critical number of components $N_c$. Again the value of $N_c$ is suggested 
to be
\begin{align}
  2.2 \lesssim N_{c} \lesssim 2.25
\end{align}

Table \REF{MM.6} combined with the investigation in polar coordinates 
for the $N=2$-model, to be described below, suggests that the
precision of $\mu_2$ at $K=18$ is three digits. 

\begin{table}[htbp]
  \begin{center}
    \begin{tabular}[c]{|l|l|l|l|}\hline
\multicolumn{1}{|c|}{$K$} & \multicolumn{1}{c|}{$\mu^{(1)}$} & 
\multicolumn{1}{c|}{$\mu^{(2)}$} & \multicolumn{1}{c|}{$\mu^{(3)}$} \\
\hline 
15  & 1.488     & 1.026      & 0.821\\
16  & 1.4557    & 1.0224     & 0.8399\\
18  & 1.43377   & 1.01889    & 0.85407\\
19  & 1.430341  & 1.018439   & 0.856580\\
\hline
    \end{tabular}
    \caption{Cubic spectrum as function of $K$}
    \label{MM.6}
  \end{center}
\end{table}

The calculation with $K=19$ was only performed for the case $N=2$ 
and $D=3$, as it required about 22 hours CPU time on an IBM RS6000 
workstation, $K=18$ required around 10 hours on the same system.

As the value $\mu_2$ decreases with increasing $K$, we expect 
the critical value $N_c$ to be given an upper bound by the value 
$N_c < 2.25$.

To find a cubic fixed point by Newton iteration, one has start from a
sufficiently close initial guess. We refrain from presenting a 
lengthy table of fixed point couplings. Instead, we present a sufficiently
accurate initial guess from which anyone can reconstruct the cubic 
fixed point. (Later we will discuss the fixed point couplings in the 
special case when $N=2$). Our search strategy was the following. 
We started at two ends of the interval of interest, at $N=2$ and at $N=3$. 
First we looked for the $O(N)$ fixed point, using the start values
\begin{align*}
& Z_{(0,0,0,0,\dots)}=0.4\\
& Z_{(1,0,0,0,\dots)}=0.06\\
& Z_{(2,0,0,0,\dots)}=0.001\\
& Z_{(n,0,0,0,\dots)}=0  ~~\forall n>2.
\end{align*}
Within our scheme (no rescaling on the $Z_{\underline{n}}$) this initial
guess turned out to be sufficient to converge to the the $O(N)$ fixed 
point at both values of $N$. To find the cubic fixed point, we took the 
$O(N)$ fixed point and changed the following coefficients:
\begin{align*}
&  Z_{(0,0,0,0,\dots)}=0.64\\
&  Z_{(0,1,0,0,\dots)}=0.007\\
&  Z_{(0,0,1,0,\dots)}=0.00011\\
\end{align*}
for the case $N=2$ and 
\begin{align*}
&  Z_{(0,0,0,0,\dots)}=0.45\\
&  Z_{(0,1,0,0,\dots)}=0.007\\
&  Z_{(0,0,1,0,\dots)}=0.00011\\
\end{align*}
for $N=3$. Having found the cubic fixed point at these two values of $N$ 
we stepped towards the critical $N_c$ from both sides, using the last 
fixed point as starting vector for the next iteration.

\def\thefootnote{\fnsymbol{footnote}}

\subsection{$\boldsymbol{N=2}$ Using Polar Coordinates}

For fixed $N$, it seems natural to introduce spherical 
coordinates. For $N=2$, we have that
\begin{equation}
Z(\phi)=
Z(r,\varphi)= \sum_{m \geq 0} Z_m(r)\, \cos(m \varphi) \, . 
\end{equation}
The cubic symmetry can be implemented by requiring that
the coefficients $Z_m(r)$ be zero for all $m$ which are 
not integer multiples of 4.
The result $Z'$ of a hierarchical RG transformation 
applied to $Z$ can be written in the form 
\begin{equation}\label{RGN2}
Z'(\phi)= {\rm e}^{-\frac12\beta^2 \phi^2} 
\int d^2 \zeta \, {\rm e}^{-\frac12 \zeta^2} \, 
{\rm e}^{\beta \phi \zeta} \, 
Z^2(\zeta) \, . 
\end{equation}
We introduce polar coordinates for $\zeta$ and $\phi$ through 
\begin{equation}
\zeta = r(\cos\varphi,\sin\varphi) \, , \quad 
\phi  = R(\cos\theta,\sin\theta)  \, , 
\end{equation}
and expand the square of $Z$ 
\begin{equation}
Z^2(\zeta)= \sum_{m \geq 0} \left(Z^2 \right)_m(r)\, \cos(m \zeta) \, .
\end{equation}
The $(Z^2)_m$ are related to the $Z_m$ through 
\begin{equation}\label{SQN2}
(Z^2)_m = \frac12 \sum_{m_1} \sum_{m_2}
Z_{m_1} \, Z_{m_2}
( \delta_{m,m_1+m_2} + \delta_{m,|m_1-m_2|} ) \, .
\end{equation}
The angle integration in eq.~(\ref{RGN2})
can be performed, resulting in 
\begin{equation}
Z'(R,\theta) = \sum_{m} Z'_m(R) \, \cos(m\theta) \, , 
\end{equation}
with 
\begin{equation}
Z'_m(R)= {\rm e}^{-\frac12 \beta^2 R^2} 
\int_0^{\infty} dr \, r \, {\rm e}^{-\frac12 r^2} \, 
I_m(\beta r R) \, (Z^2)_m(r) \, . 
\end{equation}
Here, $I_m$ denotes the modified Besselfunction of order $m$.
The next step is an expansion of the $Z_m$ in a power series of $r^2$, 
\begin{equation} 
Z_m = \sum_l Z_{ml} \, r^{2l} \, . 
\end{equation} 
Expanding the Bessel function and performing the integration
over $r$ yields the following relation (for $m$ even): 
\begin{equation} 
Z'_{ml'} = 
\sum_l 
C_{l'l}^m  \, (Z^2)_{ml}  \, ,
\end{equation} 
with 
\begin{equation}
C_{l'l}^m = 2^{l-l'} \, \beta^{2 l'} \, 
\sum_{j=0}^{l'-m/2} (-1)^j 
\frac{ (l+l'-j)!} {(l'-\frac{m}2 - j)! \, 
(l'+\frac{m}2 - j)! \, j! } \, . 
\end{equation}
A combination of the transformation with the square operation
eq.~(\ref{SQN2}) yields the structure coefficients of the complete
hierarchical RG transformation for $N=2$.  The matrix representing the
linearized transformation is then easy to compute.

\begin{figure}
\begin{center}
\includegraphics[width=13cm]{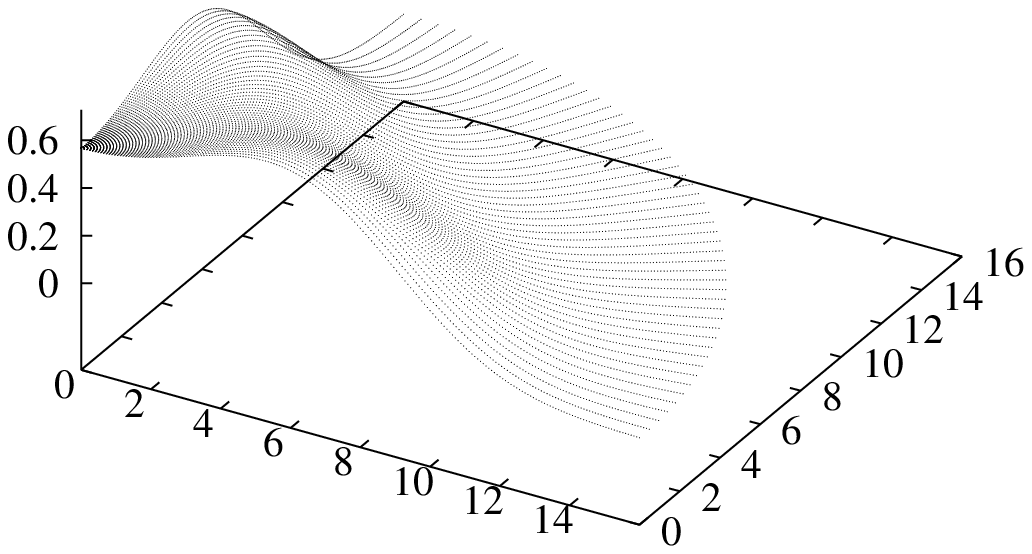}
\parbox[t]{.85\textwidth}
 {
 \caption[boltz]
 {\label{boltz}
Cubical fixed point Boltzmann factor $Z^*$ for $N=2$, 
as function of the two spin components $\phi_1$ and $\phi_2$. 
}}
\end{center}
\end{figure}

A Newton solver was used to find the cubic fixed point. It is known
\cite{GauIsi}
that for $N=2$ the cubical fixed point can be exactly mapped (by a
rotation of the spin vector with an angle $\pi/4$) to the product of
two independent Ising ($N=1$) fixed points.  Our fixed point, plotted
in figure~\ref{boltz}, is indeed of this type.

We determined the eigenvalues of the linearized transformation 
at that fixed point. Some care was devoted again to 
study the truncation effects. Table~\ref{eig2} shows
the six leading eigenvalues at the cubical fixed points for 
three different truncation orders. 
To the given precision, it is sufficient to include 
$K=32$ powers of $r^2$ in the ansatz. 
Furthermore, it is sufficient to include 
``angular momenta'' $m$ with $m \leq m_{\rm max} = 4$. 
Going to $m_{\rm max}=5$ did not change the results
for the exponents at all. 
From the factorization of the cubical fixed point into 
two Ising fixed points it follows that the spectrum 
can be built from the Ising spectrum through 
\begin{equation}
\lambda_{i,\rm cub} = \frac12 \, \lambda_{j, \rm Isi} \, 
\lambda_{k, \rm Isi}  \, . 
\end{equation}
The leading Ising eigenvalue is  1.4271725 
\cite{Pinn/Pordt/Wieczerkowski:1994}. Its square divided by two 
is 1.0184107, in nice agreement with the result for the 
subleading eigenvalue of the cubic fixed point. The whole 
Ising spectrum is part of the cubic spectrum as follows 
when one of the Ising eigenvalues is taken equal to the 
volume eigenvalue two.

\begin{table}
\small 
\begin{center}
\begin{tabular}{|l|l|l|l|l|l|c|}
\hline
   $k$ & $K=30$ 
       & $K=26$ 
       & $K=20$ 
       & $K=19$ 
       & $K=18$ 
       & $\mu$ \\ 
\hline 
   1   & 1.42717   & 1.42718  & 1.42867  & 1.43034  & 1.43377  &1.4337\\ 
   2   & 1.01841   & 1.01840  & 1.01831  & 1.01844  & 1.01889  &1.0188\\
   3   & 0.859411  & 0.859393 & 0.857925 & 0.856580 & 0.854073 &0.8540\\
   4   & 0.613255  & 0.613058 & 0.605915 & 0.602012 & 0.596873 &0.5968\\
   5   & 0.479634  & 0.479563 & 0.476318 & 0.473871 & 0.469780 &0.4887\\
   6   & 0.369252  & 0.368630 & 0.360018 & 0.357220 & 0.353718 &0.4697\\
\hline 
\end{tabular}
\caption[eig2]
{\label{eig2}
The first 6 eigenvalues of the spectrum at the cubical fixed 
point, $N=2$, for three different truncation orders.
$m_{\rm max}=4$ in all cases. 
For comparison the last column shows the results
from the calculation with the over determined basis,
truncation order $K=18$, cf.\ table~\REF{MM.5}. 
}                        
\end{center}
\end{table}

The comparison reveals that the spectrum in the polar coordinate
representation and the spectrum in the overdetermined representation
agree. Obviously, $K=18$ does not bring about a very high degree of
accuracy for the cubic invariant fixed point in the overdetermined
representation. However, larger values of $K$ 
would have required too much computer time.
Furthermore, we learn that the overdetermined
representation generates new spurious eigenvalues. They should
converge to the true ones in the no-truncation limit. The first such
spurious eigenvalue is however a subleading (or irrelevant) one.

It might be interesting to have a look at the fixed point couplings 
themselves. They are presented for $(K,m_{\rm max})=(30,4)$ in 
table~\ref{fix2}.

\begin{table}
\small 
\begin{center}
\begin{tabular}{|r|l|l|l|l|l|}\hline
  $l$&   $Z_{0l}$ & $Z_{1l}$ & $Z_{2l}$ & $Z_{3l}$ & $Z_{4l}$ \\
\hline 
   0 &  .5668E+00  &  .0000E+01  &  .0000E+01  &  .0000E+01  &  .0000E+01 \\
   1 &  .6404E--01 &  .0000E+01  &  .0000E+01  &  .0000E+01  &  .0000E+01 \\
   2 &  .3162E--02 &  .1518E--03 &  .0000E+01  &  .0000E+01  &  .0000E+01 \\
   3 &  .9355E--04 &  .1190E--04 &  .0000E+01  &  .0000E+01  &  .0000E+01 \\
   4 &  .1902E--05 &  .4309E--06 &  .3827E--08 &  .0000E+01  &  .0000E+01 \\
   5 &  .2877E--07 &  .9721E--08 &  .2351E--09 &  .0000E+01  &  .0000E+01 \\
   6 &  .3409E--09 &  .1550E--09 &  .6851E--11 &  .3149E--13 &  .0000E+01 \\
   7 &  .3280E--11 &  .1877E--11 &  .1269E--12 &  .1624E--14 &  .0000E+01 \\
   8 &  .2633E--13 &  .1807E--13 &  .1689E--14 &  .4026E--16 &  .1135E--18 \\
   9 &  .1799E--15 &  .1430E--15 &  .1728E--16 &  .6414E--18 &  .5103E--20 \\
  10 &  .1064E--17 &  .9518E--18 &  .1420E--18 &  .7407E--20 &  .1111E--21 \\
  11 &  .5515E--20 &  .5436E--20 &  .9673E--21 &  .6629E--22 &  .1567E--23 \\
  12 &  .2532E--22 &  .2703E--22 &  .5586E--23 &  .4798E--24 &  .1610E--25 \\
  13 &  .1039E--24 &  .1185E--24 &  .2783E--25 &  .2895E--26 &  .1290E--27 \\
  14 &  .3836E--27 &  .4624E--27 &  .1214E--27 &  .1488E--28 &  .8391E--30 \\
  15 &  .1283E--29 &  .1620E--29 &  .4688E--30 &  .6628E--31 &  .4568E--32 \\
  16 &  .3911E--32 &  .5135E--32 &  .1618E--32 &  .2593E--33 &  .2125E--34 \\
  17 &  .1090E--34 &  .1479E--34 &  .5024E--35 &  .9000E--36 &  .8583E--37 \\
  18 &  .2786E--37 &  .3890E--37 &  .1412E--37 &  .2791E--38 &  .3042E--39 \\
  19 &  .6533E--40 &  .9346E--40 &  .3596E--40 &  .7760E--41 &  .9517E--42 \\
  20 &  .1401E--42 &  .2046E--42 &  .8292E--43 &  .1934E--43 &  .2633E--44 \\
  21 &  .2728E--45 &  .4055E--45 &  .1719E--45 &  .4294E--46 &  .6415E--47 \\
  22 &  .4759E--48 &  .7182E--48 &  .3167E--48 &  .8407E--49 &  .1363E--49 \\
  23 &  .7311E--51 &  .1117E--50 &  .5100E--51 &  .1428E--51 &  .2490E--52 \\
  24 &  .9663E--54 &  .1493E--53 &  .7021E--54 &  .2061E--54 &  .3832E--55 \\
  25 &  .1068E--56 &  .1666E--56 &  .8042E--57 &  .2461E--57 &  .4847E--58 \\
  26 &  .9536E--60 &  .1499E--59 &  .7407E--60 &  .2353E--60 &  .4879E--61 \\
  27 &  .6560E--63 &  .1039E--62 &  .5239E--63 &  .1721E--63 &  .3738E--64 \\
  28 &  .3246E--66 &  .5174E--66 &  .2658E--66 &  .8996E--67 &  .2039E--67 \\
  29 &  .1023E--69 &  .1641E--69 &  .8570E--70 &  .2982E--70 &  .7026E--71 \\
  30 &  .1539E--73 &  .2479E--73 &  .1315E--73 &  .4694E--74 &  .1146E--74 \\
\hline
\end{tabular}
\parbox[t]{.85\textwidth}
 {
 \caption[fix2]
 {\label{fix2}
\small
Coupling constants of the cubical fixed point, for 
$K= 30$ and $m_{\rm max}=4$.
}}                        
\end{center}
\end{table}

\section{Conclusions}

In the framework of the hierarchical renormalization group, we have
studied the stability of both the $O(N)$ symmetric and the cubical
fixed point for $D=3$, in the range between $N=2$ and $N=3$. No
problems arised when investigating the stability of the $O(N)$
symmetric fixed point. For the cubical fixed point, however, the
extension of a suitable basis to non-integer $N$ turns out to be
nontrivial. We solved this by using an over-complete set of functions.
The big number of couplings that had to be used in this approach
required quite an effort to solve the fixed point equations.
Furthermore, it might be considered as a problem that the continuation
from integer $N$ to the real domain is by no means unique.  The fact,
however, that we find consistent values of $N_c$ both at the $O(N)$
and the cubical fixed point seems to indicate that the chosen basis is
a natural one.  One further comment is in order: While in the full
model the value of $N_c$ is very close to 3, the hierarchical $O(N)$
fixed point in three dimensions becomes unstable with respect to cubic
perturbations already at $N_c= 2.219$. It has been observed in other
contexts that the dependence of certain quantities on the number of
spin components is shifted towards smaller values of $N$ in the
hierarchical approximation. For instance, the two dimensional hierarchical 
non-linear $\sigma$-model is asymptotically free for $N>1$ as 
opposed to $N>2$ in the full setting 
\cite{Gawedzki/Kupiainen:1986,Pordt/Reisz:1991}. 
This is due to the absence 
of wave function renormalization in the hierarchical approximation.

Supplementing the present work by a high order $\varepsilon$-expansion
(which is certainly feasible) would be interesting and useful. 

\section*{Acknowledgements}

We would like to thank Andreas Pordt and Johannes G\"ottker-Schnetmann
for helpful discussions on the $N$-component hierarchical model.

\newpage
\section*{Appendix: A Basis for the $\boldsymbol{N=2}$ Case}

The polynomials \REF{5.2} do not form a basis for 
cubic invariant functions. There are linear dependencies. The simplest 
relation among these is given by
\begin{align}
  P_{(0,2,0,0)} = 2P_{(0,0,0,1)} - 2P_{(2,1,0,0)} + P_{(4,0,0,0)},
\label{lincomb}
\end{align}
where the indices are to be understood in the sense of \REF{5.1} with all 
$n_i$ for $i > 4$ are equal to zero.

For the case $N=2$ a basis for polynomials of type \REF{5.2} is given by
\begin{align}
  P_{(n_o,n_c)}(\phi) := \left (\sum_{a=1}^2 \phi_a^2\right )^{n_o} 
\sum_{a=1}^2 \phi^{4n_c}.
\label{n=2basis}
\end{align}
The relation with \REF{5.1} is that \REF{n=2basis} equals 
$P_{\underline{m}}$ with $m_{1}=n_{o}$, $m_{4n_{c}}=1$, and all others
are zero.

This basis has the following complications. A product of two such 
polynomials is no more the polynomial to the index given by the sum of 
the  indices of the factors. In general, a product of two basis elements 
is a linear combination
\begin{align}
   P_{(n_o,n_c)}(\phi)P_{(m_o,m_c)}(\phi) = \sum_{\substack{(k_o,k_c) \\ 
k_0=n_o+m_o\\k_c=n_c+m_c}}C^{(n_o,n_c),(m_o,m_c)}_{(k_o,k_c)}  
P_{(k_o,k_c)}(\phi).
\label{proddecompo}
\end{align}
We do not have closed expressions for the coefficients 
$C^{(n_o,n_c),(m_o,m_c)}_{(k_o,k_c)}$. We tabulated those we needed
in our programs. The table was generated using computer algebra.

The Gaussian integration of polynomials of type \REF{n=2basis} is given by
\begin{align}
\begin{aligned}
   \int\MRM{d}\mu_{\gamma}(\zeta)\;P_{(n_o,n_c)}(\psi+\zeta) = & 
\sum_{k_o=0}^{n_o} \sum_{k_c=0}^{n_c} I_{k_o,k_c}^{n_o,n_c}(N,\gamma) 
P_{(k_o,k_c)}(\psi) + \\
& \sum_{k_o=0}^{n_o} \sum_{k_c=0}^{n_c} I_{k_o,k_c+\frac{1}{2}}^{n_o,n_c}
(N,\gamma) P_{(k_o,k_c+\frac{1}{2})}(\psi)
\label{n=2gaussint}
\end{aligned}
\end{align}
with
\begin{align}
  I_{k_o,k_c}^{n_o,n_c}(N,\gamma) & = G_{2n_c,2k_c}(1,\gamma) 
G_{n_o,k_o}(4n_c+4k_c+N,\gamma),
\end{align}
where $G_{n,m}(N,\gamma)$ is as in \REF{2.4}. 
The formula \REF{n=2gaussint} does not yet provide a formula for the 
Gaussian integration in the basis \REF{n=2basis}, as the second term 
is not yet expanded in this basis. 
Its expansion involves again rather complicated structure coefficients
which we determined by means of computer algebra.

We used this representation to check our results with the overdetermined
representation. This polynomial basis turned out to a less appropriate 
representation than the following one with polar coordinates. Both 
suffer the drawback that they apply to $N=2$ only, and cannot be 
continued to any $N$. (They can be continued but do not coincide with
the desired models at integer $N$.)

\end{document}